\documentclass[12pt,iop]{emulateapj}

\usepackage{psfig}
\usepackage{apjfonts}
\usepackage{mathrsfs}

\begin{document}

\title{Constraining the true nature of an exotic binary in the core of NGC~6624.}
  \footnotetext[1]{Based on observations
    collected with the NASA/ESA {\it HST}, obtained at the Space
    Telescope Science Institute, which is operated by AURA, Inc.,
    under NASA contract NAS5-26555. }

\author{Emanuele Dalessandro\altaffilmark{2},
Cristina Pallanca\altaffilmark{2},
Francesco R. Ferraro\altaffilmark{2},  
Barbara Lanzoni\altaffilmark{2},
Claudia Castiglione\altaffilmark{2},
Cristian Vignali\altaffilmark{2},
Giuliana Fiorentino\altaffilmark{2}
}
\footnotetext[2]{Dipartimento di Fisica e Astronomia Universit\`a di
Bologna, viale Berti Pichat 6/2, I--40127 Bologna, Italy;
emanuele.dalessandr2@unibo.it},

\begin{abstract} 
We report on the identification of the optical counterpart to Star1, the exotic object serendipitously 
discovered by Deutsch et al. 
in the core of the Galactic globular cluster NGC~6624.
Star1 has been classified by Deutsch et al. 
as either a quiescent Cataclysmic Variable or a low-mass X-ray binary. 
Deutsch et al.
proposed StarA as possible optical counterpart to this object.  We
used high-resolution images obtained with the \textit{Hubble Space
Telescope} to perform  a variability analysis of the stars close to the
nominal position of Star1.  While no variability was
detected  for StarA, we found another star, here named COM$\textunderscore$Star1, showing  a
clear sinusoidal light modulation with amplitude $\Delta m_{F435W}\sim0.7$mag  
and  orbital period of $P_{orb}\sim98$min. The shape of the light curve is likely caused by strong irradiation by the 
primary heating one hemisphere of the companion, thus suggesting a quite hot primary.

\end{abstract}

\section{Introduction}

The high stellar densities typical of globular clusters (GCs) make stellar interactions very likely
events. Therefore it is expected that stellar evolution is strongly affected by the environment in
these systems and that GCs are efficient furnaces of exotic populations, i.e. systems thought to result from
the evolution of various kinds of binary systems originated or hardened by stellar interactions
(Clark 1975; Hills \& Day 1976; Bailyin 1992; Ferraro et al 2001, 2009; Ivanova et al. 2008). Indeed 
low-mass X-ray binaries (LMXBs), cataclysmic variables (CVs), millisecond pulsars and blue straggler stars
are preferentially found in GCs (Bailyn 1995; Paresce et al 1992; Ransom et al 2005; Pooley \& Hut
2006; Ferraro et al 2006, 2012).
Within this vast zoology, CVs and LMXBs deserve particular attention (for example, Ivanova et al. 2008, 
Knigge 2012 for a review).
 
CVs are binary systems composed of an
accreting White Dwarf (WD) and, typically, an unevolved companion in its core-hydrogen-burning phase.
They are relatively abundant  and since their variability is easy to detect, they 
represent an ideal benchmark to
interpret other classes of close binary systems.  CVs provide the opportunity to study binary systems
with long-term stable mass-transfer  and their analysis is an open window on the physical mechanisms
driving the evolution of many other types of contact binaries.   CVs are believed to be progenitors
of Type Ia supernovae; moreover there is an increasing evidence that some aspects of the 
accretion process taking place in CVs are similar to those observed in binary systems involving
neutron stars (NSs) or black holes (BHs; see for example Long \& Knigge 2002  and Kording et al. 2008).\\  
Depending on the intensity of the WD magnetic field, the accretion of matter from the secondary star
onto  the primary can occur either via an accretion disk in non-magnetic CVs,  or  across the
magnetic field lines in the case of polar CVs.
CVs have typical separations $a \sim R_{\odot}$ and orbital periods in the range  $P_{\rm orb} = 80
- 500$ min. Changes in the period of CVs are linked to changes in the
structure of the donor and to the physical mechanism that drives the accretion of matter onto the
WD.

LMXBs are Roche-lobe filling binary systems consisting of a NS or a BH 
accreting mass from a low-mass star. They are believed to be the progenitors of millisecond pulsars. 
In GCs most of NSs or BHs lose their primordial 
binary companion when they form. Then, these compact objects can capture a new companion 
via binary exchange, physical collisions with giants or tidal capture.
Tidal capture is a poorly efficient formation mechanism in GCs (Ivanova et al. 2008), physical
collisions typically lead to the formation of NS-WD LMXBs, the so called Ultra-compact XBs (UCXBs; Ivanova et
al. 2005), while binary exchanges may lead to the formation 
of binary systems with any kind of companions. LMXBs span a vast range of orbital periods 
from a few minutes (see the case of 4U 1820-30 in NGC6624 -- Stella et al. 1987 or M15-X2 -- Dieball et al.
2005), to several hours. They are typically characterized by soft X-ray spectra.  
 
\begin{figure*}[]
\includegraphics[width=180mm]{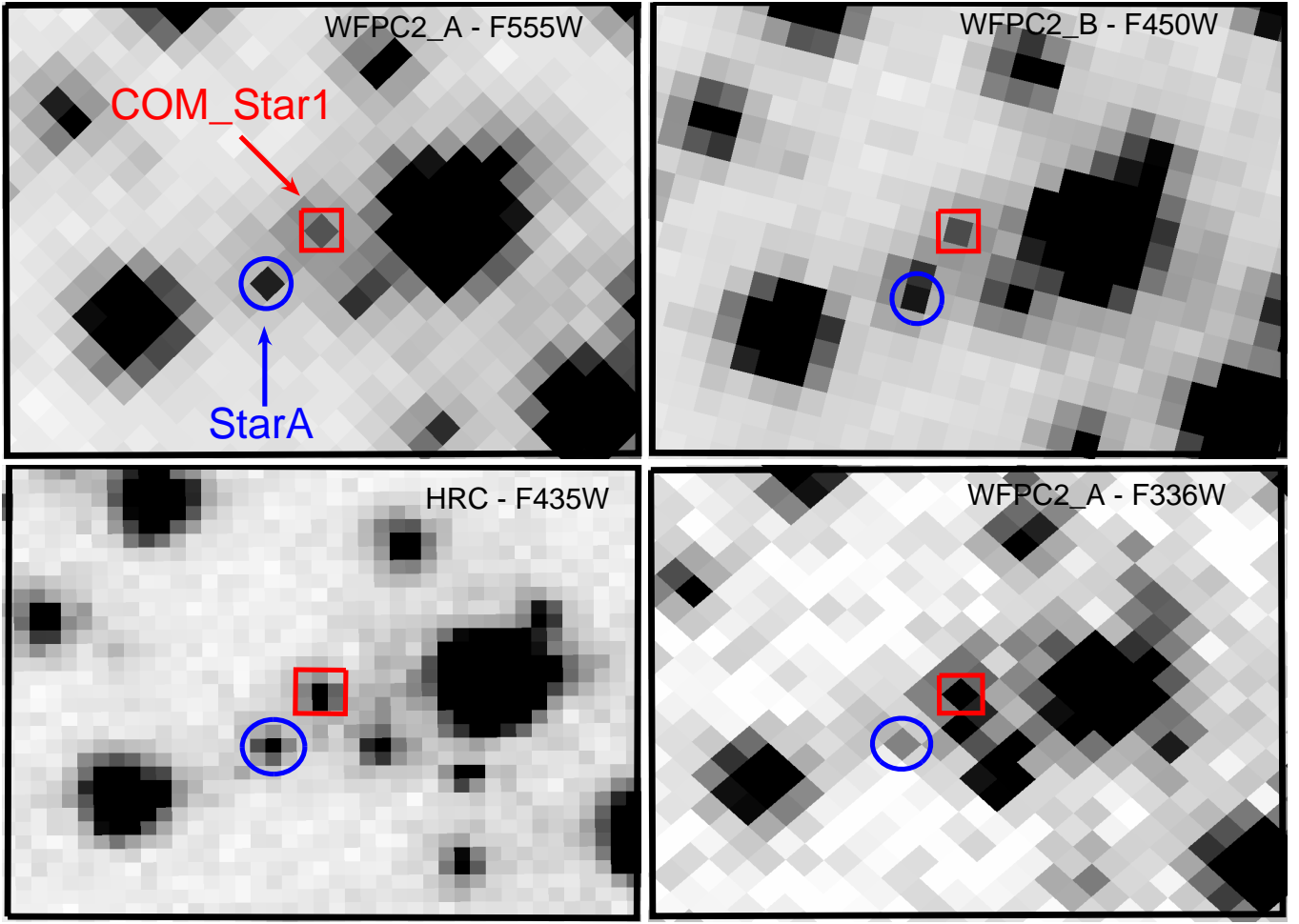}
\caption{Zoomed view around the position of StarA and  COM$\textunderscore$Star1 in four images at 
different wavelengths. The size of the images 
is about $0.4\arcsec\times0.4\arcsec$. North is up, East is left.}
\label{map}
\end{figure*}

Deutsch et al. (1999; hereafter D99) reported on the serendipitous discovery of an exotic object in the core of
NGC~6624.  With the aim of observing StarK (4U 1820-30, Stella, Priedhorsky \& White 1987; King et al. 1993; Anderson
et al. 1997), which is a UCXB, they used the
Space Telescope Imaging  Spectrograph (STIS) on board HST with the long-slit configuration ($0.5\arcsec \times
23\arcsec$).  At a few arcsec from StarK and within the field of view covered by the slit, they found two
additional stars.  One of the two objects (Star1), located at $4.66\arcsec$ from StarK, showed strong and broad
emission  lines and a very weak continuum.  This object has been
classified as a classical CV in a quiescent state (Wu et al. 1992), 
although the observed spectrum did not allow to unambiguously exclude the possibility that Star1 is a quiescent
LMXB. \\ 
On the basis of its position in the HST Faint Object Camera and Wide Field Planetary Camera 2 (WFPC2)
images, D99 identified a faint star (see their Table~1), named StarA, as the possible optical counterpart to
Star1. This identification was mostly based on the relative positions between StarK and StarA, defined by
the World Coordinate System (WCS) of the HST headers. Given the internal uncertainties ($\sim0.5\arcsec$), the authors were
not able to discard other possible counterparts.  
In order to constrain the preliminary D99 identification, we have performed a detailed 
analysis of a set of optical
and UV HST images of the core of NGC~6624. We have found that StarA does not show any evidence of
variability, while a close object (at $0.15\arcsec$) is characterized by a clear light modulation. 
Here we propose that this object is the real counterpart to Star1, 
we determine its period and discuss the possible nature of the system.

\section{Observations and Data Analysis}

For the present analysis we used a combination of HST WFPC2 and 
Advanced Camera for Survey - High
Resolution Channel (HRC) images.
The WFPC2 images have been obtained trough three different proposals and we will group them here
into two samples.
The first sample  (WFPC2$\textunderscore$A; Prop ID:11975 - PI: Ferraro) consists of 13 images obtained 
in four different filters: F555W (three images with $t_{exp}=100$ sec), 
F336W (three with $t_{exp}=700$ sec), 
F255W (four with $t_{exp}=1200$ sec) and F170W (two with $t_{exp}=1200$ sec and one with $t_{exp}=1300$ sec).
The second sample (WFPC2$\textunderscore$B; Prop ID: 10841, 11988 - PI: Chandar) consists of a total of 40 images obtained with the
F450W filter with $t_{exp}$ ranging between 300 sec and 400 sec.
The HRC data consist of 20 F435W images with $t_{exp}=200$ sec each (Prop ID: 10401 - PI: Chandar). 
HRC images have been corrected for 
Pixel Area Map by using the files available 
in the HST web site. 
The HRC pointings fall almost completely within the Planetary Camera (PC) field of view which
is approximately located on the cluster center. For this work we used only the PC and HRC
images. \\
For all datasets, the photometric reduction has been performed by using DAOPHOTII (Stetson 1987).
Tens of bright and isolated stars have been selected in each frame to model the Point Spread Function.
A first star list has been obtained for each image by fitting independently all
the star-like sources detected at the $3\sigma$ level from the local background.
In order to exploit the high spatial
resolution of the HRC images, we forced (for details see Dalessandro et al. 2011, 2013) 
the positions of stars identified in at least 
10 HRC images to be fitted in all the WFPC2-PC images by using ALLFRAME (Stetson 1994).\\
WFPC2 instrumental magnitudes have been corrected for Charge Transfer Efficiency effect by using
prescriptions and equations reported by Dolphin (2000). 
Magnitudes have been then reported to the VEGAMAG photometric system by following
Holtzmann et al. (1995) and zeropoints provided in the dedicated HST web-page. 
Also the instrumental HRC $m_{F435W}$
magnitudes have been calibrated to the same photometric system by using zeropoints 
and equations by Sirianni et al. (2005). 
We then reported the WFPC2$\textunderscore$B  $m_{F450W}$ measures to the HRC $m_{F435W}$ system by
applying small zeropoints estimated by using the Main Sequence (MS) stars in common between the HRC 
and WFPC2$\textunderscore$B samples.\\
For each star, different magnitude estimates have been homogenized (see Ferraro et al 1991,1992) and
their weighted mean and standard deviation have been 
finally adopted as the star magnitude and its photometric error.
Instrumental coordinates have been roto-trasled to the absolute coordinates system by using the 
catalog of astrometric standards {\it Guide Star Catalog 2.3} and the cross-correlation software
CataXcorr. At the end of the procedure, the typical accuracy of the astrometric solution is of the order of
$0.1\arcsec$.

\begin{figure}
\includegraphics[width=90mm]{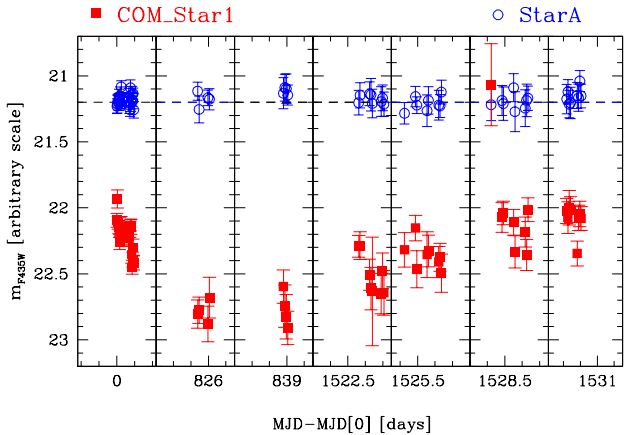}
\caption{Light curve of
  COM$\textunderscore$Star1 and StarA in seven different time intervals.  Each
  time bin spans about 0.45 days. MJD[0] corresponds to the
  observation time of the oldest image in our sample (see Section~3).}
\label{lc}
\end{figure}

\section{The optical counterpart to Star1}

We identified StarA (see Figure~1) in our catalog by using its coordinates in the HST WCS 
reported by D99 and the finding charts in their Figure~3.\\ With the aim of inferring the orbital
period and the nature of Star1, we performed a detailed variability analysis of StarA.  For
this purpose we used the HRC and WFPC2$\textunderscore$B samples, since they are the data-sets with a number of
exposures large enough for this kind of analysis. In this way we can use 60 homogeneous measures
(see Section~2) covering a total baseline larger than four years (Figure~2) starting at $MJD[0]= 53429.89947762$ days, which
corresponds to the oldest image of our sample.\\
We found that StarA does not show  any
evidence for flux modulation (Figure~2). For this reason, we extended the
variability analysis to all stars lying within $\sim 1\arcsec$ from
StarA and we detected a clear luminosity variation up to about $\Delta m_{F435W}=0.7$ mag  (Figure~2)
in a faint star (hereafter named COM$\textunderscore$Star1)
located $\sim 0.15\arcsec$ NW from StarA (Figure~1). 
Although D99 considered the possibility that this star might be the counterpart of Star1 
(see their Section~2.2), their analysis was focused only on StarA. 
It is also important to note that the position of COM$\textunderscore$Star1 is fully compatible with that of 
the STIS spectrum observed by D99.\\ 
Both stars are relatively faint objects and they do not
show any UV emission in the $F255W$ and $F170W$ frames.
\begin{figure}
\includegraphics[width=90mm]{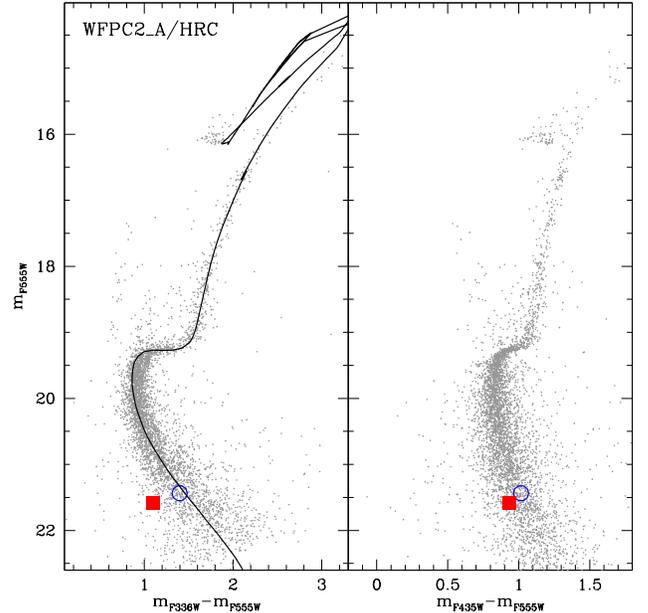}
\caption{$(m_{F555W}, m_{F336W}-m_{F555W})$ and $(m_{F555W}, m_{F435W}-m_{F555W})$ CMDs of the 
WFPC2-HRC field of view (\textit{left} and
\textit{right panel} respectively). The positions in the CMDs of StarA and  COM$\textunderscore$Star1
 are highlighted by a blue 
circle and a red square respectively. In the \textit{left panel}, we show the best fit isochrone ([Fe/H]=-0.44, $t=12$Gyr
from the BaSTi database - Pietrinferni et al. 2006). }
\label{map}
\end{figure}
This is compatible with the UV
limits reported by D99 for StarA and the detection limits of our data.\\ The positions of both
StarA and COM$\textunderscore$Star1 in the optical $(m_{F555W}, m_{F336W}-m_{F555W})$ and  $(m_{F555W},
m_{F435W}-m_{F555W})$ Color Magnitude Diagrams (CMDs) are shown in Figure~3.  StarA is likely a MS star,  its
magnitudes $m_{F336W}$, $m_{F435W}$ and $m_{F555W}$ are fully consistent with what found by
D99, once  differences in the adopted photometric systems are properly taken into account (see
their Table~1). COM$\textunderscore$Star1 is slightly bluer than StarA by $\Delta(m_{F435W}-m_{F555W})\sim0.2$
and $\Delta(m_{F336W}-m_{F555W})\sim0.4$, and is located at blue edge
of the MS in both CMDs. \\  
We analyzed the light curve of COM$\textunderscore$Star1 by using the Graphical Analyser
of Time Series  ({\rm GRATIS}) which is a private software developed at the Bologna
Observatory by P. Montegriffo. We looked for possible solutions to the observed light curve
in the period range $P\sim 30 - 700$ min, which
well contains the period distribution of known CVs and LMXBs. 
By following Saha \& Hoessel (1990), we estimated that, in this
period range and with the available dataset, the probability of deriving the correct period 
of a variable star is always larger than $\sim80\%$.\\
In Figure~4 we show the Lomb periodograms (Lomb 1976) for both StarA and COM$\textunderscore$Star1.
While StarA does not show any significant peak, COM$\textunderscore$Star1 shows a prominent one at 
Frequency =14.67 cycles d$^{-1}$  (corresponding to $P_{\rm orb}\sim98$ min) with a confidence level
larger than $3\sigma$ (considering Poissonian noise).\\
We find that the light curve of COM$\textunderscore$Star1 is well folded by a period $P_{\rm
orb}\sim 98$ min.
The period-folded light curve is shown in Figure~5 where we adopted as reference time $MJD[0]$.
The time baseline 
covered by the adopted data-set corresponds to
about 24500 COM$\textunderscore$Star1 cycles, during which the estimated period seems to be stable. 
Therefore 
we conclude that the periodic signal detected in the observed light curve 
likely corresponds the orbital period of COM$\textunderscore$Star1\footnote{Since $P_{\rm orb}$ is 
very close to the HST orbital period ($P_{\rm HST}\sim96$min), 
we adopted the approach described 
by Zurek et al. (2009) finding a quite small probability ($\sim1\%$) that $P_{\rm orb}$ is an alias of $P_{\rm HST}$.}.\\
With the adopted data-set the orbital period of COM$\textunderscore$Star1 is sampled approximately 8-9 times.
The light curve shows a quite regular, almost sinusoidal behavior. 
The flux modulation of COM$\textunderscore$Star1 shows also some noisy aperiodic 
variability (flickering), 
with a few measures appearing as strong outliers to the general behavior.
In particular at $\phi\sim0.4$ we observe a single measure brighter by about 1 mag 
than the observed
luminosity at the maximum.
As recently shown by Ingram \& van der Klis (2013) and Scaringi (2013) such aperiodic variability
is likely the result of fluctuations in the mass-transfer rate at different radii.  

\vspace{0.5cm}

\begin{figure}
\includegraphics[width=90mm]{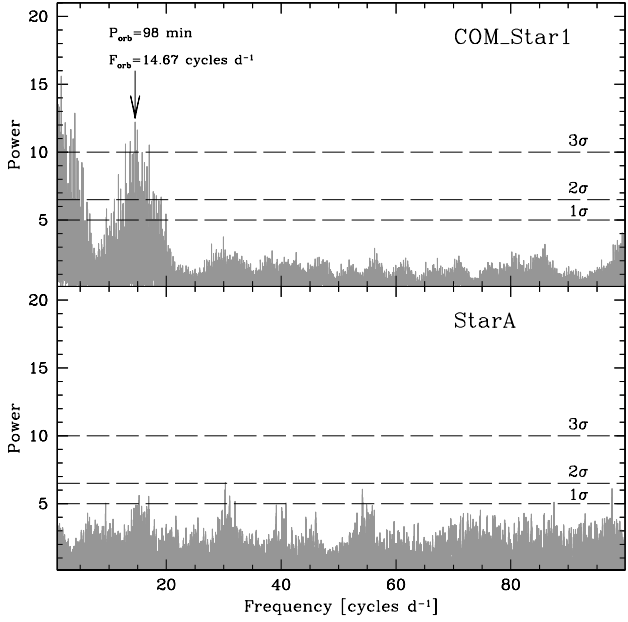}
\caption{Lomb periodograms for COM$\textunderscore$Star1 and StarA.
Dashed lines are
different confidence levels estimated by considering Poissonian noise. 
}
\label{map}
\end{figure}

\section{Results and Discussion}

If Star1 is a CV (D99), the value of the period best folding the light
curve of COM$\textunderscore$Star1 places this system approximately at
the peak of the period distribution of known CVs.

Indeed, the light curve of COM$\textunderscore$Star1 is very similar
to those observed in some of the recently confirmed polar CVs (Margon
et al. 2013), and in the well studied cases of V1974 Cygni (Nova Cygni
1992; De Young \& Schmidt 1994) and V2275 Cygni (Balman et al. 2013).
The sinusoidal shape can be interpreted as being caused by the orbital
revolution of an hemisphere of the secondary star heated by radiation
from the hot WD. In fact, as detailed by Kovetz, Prialnik \& Shara
(1998), when an outburst occurs, the WD hydrogen-rich envelope is
ejected in few months, and the WD outer layer remains at a temperature
of the order of $\sim10^5$ K, irradiating the secondary star and
keeping high rates of mass-transfer ($\sim 10^{-8} -
10^{-9} M_{\odot} yr^{-1}$) for some centuries.  Consistently, V1974 and
V2275 Cygni have been observed just 1-2 years after their nova
outburst phase and show sinusoidal light curves.  For the case of
Star1, we do not have knowledge of its most recent outburst and we do
not observe any appreciable difference in the luminosity of
COM$\textunderscore$Star1 between the data presented by D99 and those
shown in this work. However, the possibility of a heating mechanism
also in the case of COM$\textunderscore$Star1 is suggested both by the
sinusoidal shape of its light curve, and by the observed
($m_{F336W}-m_{F435W}$) color variation as a function of the orbital phase.
In fact we estimate a temperature variation
ranging between $T\sim5000$K and $T\sim9000$K, thus yielding a lower
limit of about $\Delta T\sim4000K$\footnote{Note that this estimate is based only on three measures.}.

To quantitatively investigate the possibility that the heating scenario
applies to Star1, we compared the observed flux variation, $\Delta
F_{\rm obs}\sim 5\times10^{-15}$ erg cm$^{-2}$ $s^{-1}$, as obtained
from the light curve in Figure~5 to the value expected in the case of irradiation (see
Pallanca et al. 2012 and references therein):
\begin{displaymath} 
\Delta F_{exp}(i)= \eta \sigma \frac{R^2_{\rm WD} T^4_{\rm WD}
  R^2_{\rm COM} }{a^2} \frac{i}{\pi d^2}(1-\frac{R^2_{\rm COM}}{a}).
\end{displaymath} 
Here $\eta$ is the reprocessing efficiency (which we assumed to be
$\eta=0.5$), $\sigma$ is the Stefan-Boltzman constant, $a$ is the
apparent orbital separation ($a\sim0.65 R_\odot$), $d$ is the distance
of the system (which we assumed to be the one of NGC6624: $d=8.4$ kpc,
from Valenti et al. 2007), $R_{\rm COM}$ is the radius of the
companion star (which we assumed to be equal to the Roche Lobe radius:
$R_{\rm COM}\sim0.2R_\odot$), and $i$ is the system inclination.

To determine the orbital separation (from the Kepler's third law) and
the Roche Lobe radius (by following Paczynski 1971), we first derived
the primary and secondary star parameters as follows.  We assumed a WD
mass $m_{\rm WD}\sim0.6 M_{\odot}$ typical of WDs in old stellar systems as GCs, which
yields a WD radius $R_{\rm WD}\sim10^{-2} R_{\odot}$.  Then, following
the mass-period relation ($M_2=0.11P_{hr}$; Warner 1995), the mass of
the secondary is $M_2\sim0.2M_{\odot}$.  We note that by projecting
the companion star position in the optical CMD, onto the isochrone
best fitting the main evolutionary sequences of NGC 6624\footnote{We adopted E(B-V)=0.28 and $(m-M)_0=14.63$
from Valenti et al. (2007).} (Figure~3),
under the assumption that this object is still
behaving as a MS stars, we would obtain $M_2\sim0.7M_{\odot}$. However
in the case of a perturbed star, such a value can be easily
overestimated by a factor of two or three (see, e.g., Pallanca et
al. 2010) and we therefore assumed $M_2\sim0.2M_{\odot}$.

By using these values and the equation above, we finally found that, in
order to account for the observed flux variation $\Delta F_{\rm obs}$,
the effective temperature of the WD needs to be $T_{\rm WD}>45000$ K.
This temperature is at least two times larger than what derived from
the UV flux threshold of our observations and from the UV luminosity
inferred by D99 from the continuum fitting of the STIS spectrum, which
gives $T_{\rm WD}<20000$K.  Of course this estimate relies on the
assumption that Star1 is member of NGC~6624.  Unfortunately, D99 were
not able to measure the radial velocity of COM$\textunderscore$Star1
and we have verified (Bellini A., private communication) that with the
available HST data it is not possible to obtain a reliable proper
motion measure of COM$\textunderscore$Star1 to constrain its
membership.

\begin{figure*}[]
\includegraphics[width=170mm]{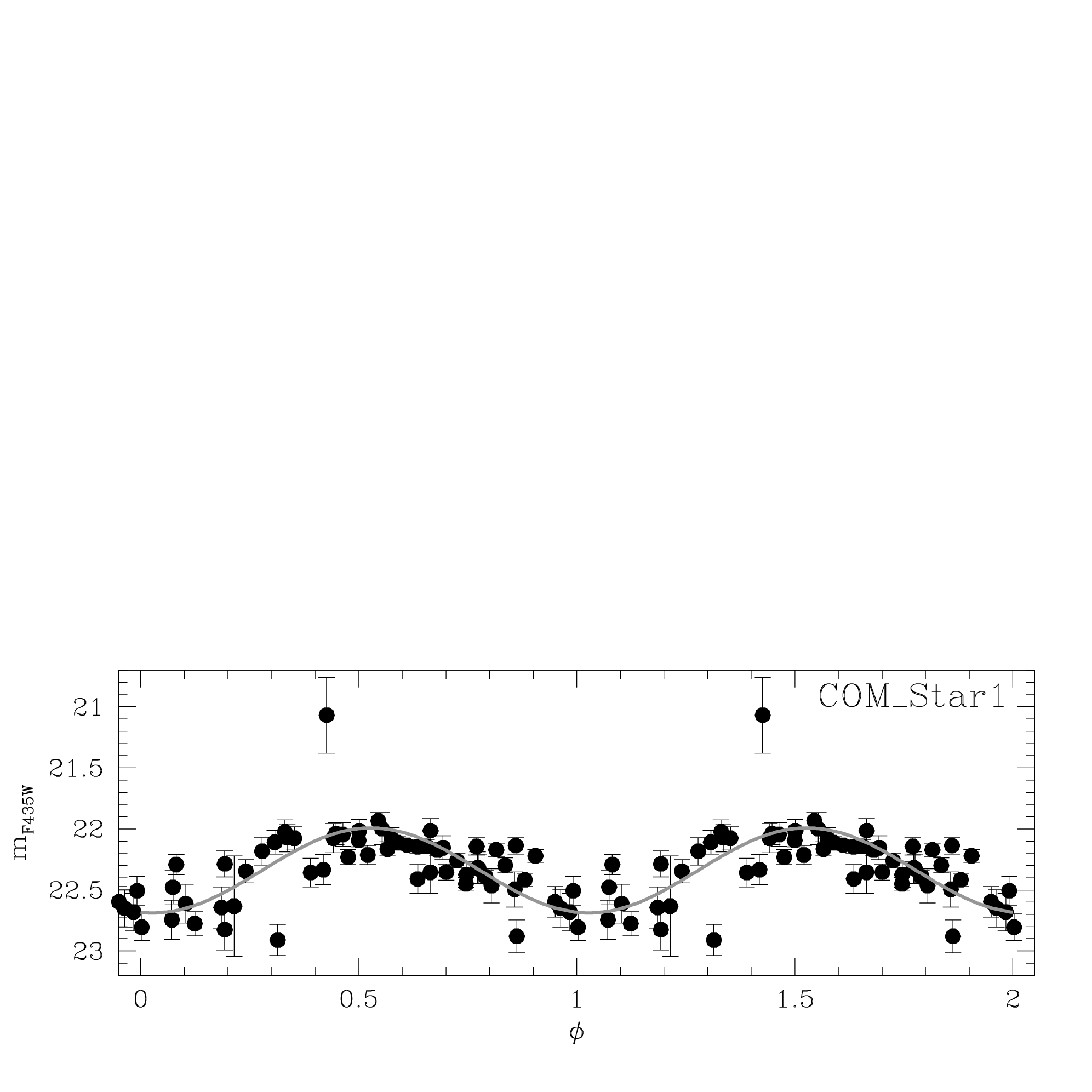}
\caption{Light curve of  COM$\textunderscore$Star1 folded with the estimated 
orbital period $P_{\rm orb}\sim98$ min. The grey line is a sinusoidal function of amplitude
$\Delta m_{F435W}\sim0.7$.}
\label{mod}
\end{figure*}

An alternative scenario to account for the observed light curve is that Star1 is a LMXB and a
NS is responsible 
for the irradiation of the secondary not degenerate star.
Following this possibility, we have analyzed the available high-resolution Chandra
X-ray spectra (Prop ID: 02400090; PI: Murray) of the core of NGC~6624. These data cannot be used to 
constrain the emission of Star1,
since the X-ray flux is totally dominated by StarK just few arcseconds apart.
D99 discussed the issue of possible
confusion of the observed STIS spectrum of Star1 with that of a quiescent LMXB (see also Grindlay 1999).
While the spectrum does not
allow a clear distinction between a CV and a LMXB, D99 preferred to conservatively classify Star1 as CV
since this class of objects is more common. 
However, the light curve presented here suggests that the secondary is undergoing an heating 
process from a quite hot primary, thus leaving open the possibility that COM$\textunderscore$Star1
is orbiting a NS.

\section{Summary and Conclusions.}

We have searched for the optical companion to Star1 by 
performing a detailed photometric variability analysis. We have found that the companion 
proposed by D99 (StarA) does not show any evidence of variability, while we have identified an object, named 
COM$\textunderscore$Star1 
showing a clear light modulation with an amplitude $\Delta m_{F435W}\sim0.7$ mag. 
The light curve of COM$\textunderscore$Star1 shows a periodic signal with $P_{\rm orb}\sim98$min,
which appears to be stable over a time interval of about four years and should therefore 
correspond to the orbital period of Star1.

The flux modulation of COM$\textunderscore$Star1 has a quite regular and sinusoidal behavior, which
might be interpreted 
as driven by irradiation by the primary star on one hemisphere of the non-degenerate secondary star.
This scenario is further supported by some hints of temperature variations as a function of the orbital phase 
inferred by the ($m_{F336W}-m_{F435W}$) color variation.
By performing simple calculations, we have constrained the temperature needed for the primary
to efficiently heat the secondary. It results that the primary should have a temperature $T>45000$K.
If Star1 is a CV, as preferentially suggested by D99, this temperature lower limit is not compatible with the one 
inferred by the UV continuum observed by D99 and by the UV detection threshold of our images.
This incompatibility might push toward the possibility that Star1 is a LMXB and the secondary is heated by 
a NS. We performed the analysis of the available Chandra data, but unfortunately 
they cannot be used to constrain the X-ray emission of Star1, since it is dominated by the flux 
of the ultra-luminous X-ray source StarK located a few arcsec apart. 
 
Of course several pieces of information are still needed to fully constrain the nature of
Star1. 
A long-term monitoring with multi-wavelength synchronized observations 
is urged to derive an accurate estimate of the temperature variation on the secondary surface. 
In addition ultra-deep UV observations can provide further constrain on the nature of the
primary.
Chandra observations in soft ($\sim0.3-1$~keV) and hard ($\sim1-4.5$~keV) bands, 
in a sub-array configuration optimized to exclude as much as possible the flux of StarK, 
can be useful to detect and possibly characterize the X-ray emission of Star1 either via direct X-ray 
spectroscopy or hardness ratio analysis. \\
A multi-period monitoring of the light curve through high-resolution optical HST observations coupled 
with appropriate models for CVs and LMXBs would definitely allow a more robust analysis and 
would shed new light on the properties of this intriguing system.\\

\acknowledgements  
This research is part of the project COSMIC-LAB funded by the European Research Council 
(under contract ERC-2010-AdG-267675). The authors thank the anonymous referee for 
the important contribution to the correct interpretation of the results. 
ED warmly thanks Giuseppe Greco
for useful discussions and suggestions about the variability analysis.

{}
\end{document}